\documentclass[12pt]{article}
\usepackage{amssymb,epsfig}
\newcommand{\be}[1]{
\begin{equation}\label{#1}}
\newcommand{\ee}{\end{equation}}
\newcommand{\ba}[1]{
\begin{eqnarray}\label{#1}}
\newcommand{\ea}{\end{eqnarray}}
\newcommand{\baa}{\begin{eqnarray*}}
\newcommand{\btab}{\begin{tabular}}
\newcommand{\etab}{\end{tabular}}
\newcommand{\eaa}{\end{eqnarray*}}

\def \labeltest #1 {\label{#1}}

\def\II{\hbox{{1}\kern-.25em\hbox{l}}}

\begin{document}

\begin{titlepage}
\begin{flushright}
\begin{tabular}{l}
 \\
CYCU-HEP-10-08
\end{tabular}
\end{flushright}
\vskip1cm
\begin{center}
  {\large \bf
  Two-parton Light-cone Distribution Amplitudes
  \\}
  {\large \bf
  of Tensor Mesons
  \\}
\vspace{1cm} {\sc Hai-Yang Cheng$^\ddag$, Yuji Koike$^*$ {\rm and} Kwei-Chou Yang$^\dag$ }

\vspace*{0.1cm} {\it $^\ddag$ Institute of Physics, Academia Sinica, Taipei, Taiwan 115, Republic of China\\
and \\
C.N. Yang Institute for Theoretical Physics,
State University of New York \\Stony Brook, New York
11794}\\

\vspace*{0.1cm} {\it $^*$Department of Physics, Niigata University, Ikarashi,
Niigata 950-2181, Japan}\\

\vspace*{0.1cm}  {\it $^\dag$Department of Physics, Chung Yuan Christian
University, Chung-Li, Taiwan 320, Republic of China}\\[1.cm]
\vskip0.8cm
{\bf Abstract\\[10pt]} \parbox[t]{\textwidth}{
We present a detailed study of the two-parton light-cone distribution amplitudes for $1\,^3P_2$ nonet tensor mesons.  The asymptotic two-parton distribution amplitudes of twist-2 and twist-3 are given. The decay constants $f_T$ and $f_T^\bot$ defined by the matrices of non-local operators on the lightcone are estimated using the QCD sum rule techniques. We also study the decay constants for $f_2(1270)$ and $f_2^\prime(1525)$ based on the hypothesis of tensor meson dominance together with the data of $\Gamma(f_2\to \pi\pi)$ and $\Gamma(f'_2\to K\bar K)$ and find that the results are in accordance with the sum rule predictions.}
\vskip1cm

\end{center}

\end{titlepage}

\newpage

\section{Introduction}

~~~ In the past few years, BaBar and Belle have measured several charmless $B$ decay modes involving light tensor mesons in the final states \cite{PDG}. These decays play a complementary role, compared with $e.g.$, $B\to VV,VA,AA$ channels ($V$ is a vector and $A$ is an axial-vector meson) \cite{BenekeVV,Cheng:2008gxa},  since the tensor meson $T$ can be produced neither from the local (axial-)vector current nor from the local tensor current which is relevant only to new physics. The polarization studies for $B\to TV, TA, TT$ decays can further shed light on the underlying helicity structure of the decay mechanism, recalling that the longitudinal polarization dominance observed in the decay $B^+\to \phi K_2^*(1430)^+$ is quite different from the polarization measurement in $B\to \phi K^*$ which indicates a large fraction of transverse polarization \cite{:2008zzd}.

In the quark model, the $J^{PC}=2^{++}$ tensor meson can be modeled as a constituent quark-antiquark pair with the angular momentum $L=1$ and total spin $S=1$. The observed tensor mesons $f_2(1270)$, $f_2'(1525)$, $a_2(1320)$ and $K_2^*(1430)$ form an SU(3) $1\,^3P_2$ nonet.  The $q\bar q$ content for isodoublet and isovector tensor resonances are obvious.  \footnote{Just as the $\eta$-$\eta'$ mixing in the pseudoscalar case, the isoscalar tensor states $f_2(1270)$ and $f'_2(1525)$ also have a mixing, and their wave functions are defined by
 \begin{eqnarray}
 f_2(1270) &=&
{1\over\sqrt{2}}(f_2^u+f_2^d)\cos\theta_{f_2} + f_2^s\sin\theta_{f_2} ~, \quad
 f'_2(1525) =
{1\over\sqrt{2}}(f_2^u+f_2^d)\sin\theta_{f_2} - f_2^s\cos\theta_{f_2} ~, \nonumber
 \end{eqnarray}
with $f_2^q\equiv q\bar q$.  Since $\pi\pi$ is the dominant decay mode of $f_2(1270)$ whereas $f_2'(1525)$ decays predominantly into $K\bar K$ (see Ref.~\cite{PDG}), it is obvious that this mixing angle should be small.  More precisely, it is found that $\theta_{f_2}=7.8^\circ$ \cite{Li} and $(9\pm1)^\circ$ \cite{PDG}.  Therefore, $f_2(1270)$ is primarily a $(u\bar u+d\bar d)/\sqrt{2}$ state, while $f'_2(1525)$ is dominantly $s\bar s$.
}
Nevertheless, in full QCD field theory, the tensor meson is represented by a set of Fock states, each of which has the same quantum number as the meson. In this work, we present the study for two-parton asymptotic light-cone distribution amplitudes (LCDAs) of lowest-lying tensor mesons with quantum numbers $J^{PC}=2^{++}$ because, in the treatment of exclusive $B$ decay processes in QCD, the Fock states of the energetic meson can be further represented in terms of LCDAs. The LCDAs are governed by the special collinear subgroup $SL(2,\mathbb{R})$ of the conformal group \cite{BF2,Braun:2003rp} and can be expanded as a series of partial waves, where the rotational invariance is characterized by the conformal spin $j$ and the concept of ``collinear twist'' is
equivalent to the ``eigen-energy" in quantum mechanics.

Due to the $G$-parity of the tensor meson, according to our definition, both the chiral-even and chiral-odd two-parton LCDAs of the tensor meson are antisymmetric under the interchange of momentum fractions of the $quark$ and $anti$-$quark$ in the SU(3) limit. The asymptotic LCDAs are relevant to the first Gegenbauer moment of the leading twist distribution amplitudes, $\phi_\parallel$ and $\phi_\perp$. In analogy to the cases of axial-vector mesons \cite{Cheng:2008gxa,Yang:2005tv}, the sizable Gegenbauer term containing the first Gegenbauer moment could have a large impact on $B$ decays involving a tensor meson.

The present paper is organized as follows. In Sec.~\ref{sec:DEF} we define the LCDAs for the tensor mesons. A slightly different definition for chiral-even LCDAs is given in \cite{Braun:2000cs}. The detailed properties of  LCDAs are given in Sec. \ref{sec:PRO}. Results for the decay constants
are presented in Sec.~\ref{sec:DC}. Sec.~\ref{sec:sum} comes to our conclusion.

\section{Definition}\label{sec:DEF}

~~~ For a tensor meson, the polarization tensors $\epsilon_{(\lambda)}^{\mu\nu}$ with helicity $\lambda$ can be constructed in terms of the polarization vectors of a massive vector state moving along the $z$-axis \cite{Berger:2000wt}
\begin{eqnarray}
\varepsilon(0)^{*\mu} = (P_3,0,0,E)/m_T,
\quad
\varepsilon(\pm1)^{*\mu} = (0,\mp1,+i,0)/\sqrt{2},
\end{eqnarray}and are given by
\begin{eqnarray}
\epsilon^{\mu\nu}_{(\pm2)} &\equiv& \varepsilon(\pm1)^\mu \varepsilon(\pm1)^\nu,
\\
\epsilon^{\mu\nu}_{(\pm1)} &\equiv& \sqrt{\frac{1}{2}}
[\varepsilon(\pm1)^\mu \varepsilon(0)^\nu + \varepsilon(0)^\mu \varepsilon(\pm1)^\nu],
\\
\epsilon^{\mu\nu}_{(0)} &\equiv& \sqrt{\frac{1}{6}}
 [\varepsilon(+1)^\mu \varepsilon(-1)^\nu + \varepsilon(-1)^\mu \varepsilon(+1)^\nu]
 + \sqrt{\frac{2}{3}}  \varepsilon(0)^\mu \varepsilon(0)^\nu.
\end{eqnarray}
The polarization $\epsilon_{\mu\nu}^{(\lambda)}$ can be
decomposed in the frame formed by the two light-like vectors, $z_\mu$ and $p_\nu\equiv P_\nu - z_\nu m_T^2/(2pz)$ with $P_\nu$ and $m_T$ being the momentum and the mass of the tensor meson, respectively, and their orthogonal plane~\cite{Ball:1998sk,Ball:1998ff}. The transverse component that we use thus reads
\begin{eqnarray}\label{eq:polprojectiors}
 && \epsilon^{(\lambda)}_{\perp\, \mu\nu} z^\nu
 =\epsilon^{(\lambda)}_{\mu\nu} z^\nu-\epsilon^{(\lambda)}_{\parallel\mu\nu} z^\nu
        = \epsilon^{(\lambda)}_{\mu\nu} z^\nu -
     \frac{\epsilon^{(\lambda)}_{\alpha\nu} z^\alpha z^\nu }{p z} \left( p_\mu-\frac{m_T^2}{2 p z} \,z_\mu\right)\,.
\end{eqnarray}
The polarization tensor $\epsilon^{(\lambda)}_{\alpha\beta}$, which is symmetric and
traceless, satisfies the divergence-free condition $\epsilon^{(\lambda)}_{\alpha\beta} P^\beta=0$ and the orthonormal condition $\epsilon^{(\lambda)}_{\mu\nu}\big(\epsilon^{(\lambda') \mu\nu}\big)^*=\delta_{\lambda\lambda'}$. Therefore,
\begin{eqnarray}
 \langle T(P, \lambda)|V_\mu|0\rangle &=&
 a\epsilon^{*(\lambda)}_{\mu\nu}P^\nu+b\epsilon^{*(\lambda)\nu}_{~\nu} P_\mu=0, \\
 \langle T(P,\lambda)|A_\mu |0\rangle &=& \varepsilon_{\mu\nu\rho\sigma} P^\nu \epsilon_{(\lambda)}^{\rho\sigma *}=0,
\end{eqnarray}
and hence the tensor meson cannot be produced from the local $V-A$ current and likewise from the tensor current.
The completeness relation reads
\begin{eqnarray} \label{eq:polarization}
 \sum_\lambda \epsilon^{(\lambda)}_{\mu\nu}\left(\epsilon^{(\lambda)}_{\rho\sigma}\right)^\ast
  = \frac12 M_{\mu\rho} M_{\nu\sigma}+\frac12 M_{\mu\sigma} M_{\nu\rho}
  -\frac13 M_{\mu\nu} M_{\rho\sigma}\,,
\end{eqnarray}
where $M_{\mu\nu} = g_{\mu\nu} - P_\mu P_\nu/m_T^2$.

In what follows, we consider matrix elements of bilocal quark-antiquark operators at a light-like separation, $2z_\mu$, with $z^2=0$. In analogy with those of vector and
axial-vector mesons \cite{Ball:1998sk,Ball:1998ff,Yang:2005gk,Yang:2007zt}, we can define chiral-even
light-cone distribution amplitudes of a tensor meson:\footnote{Our LCDA $g_a$ differs from that defined by Braun and Kivel \cite{Braun:2000cs}:
\begin{eqnarray}
 \langle T(P,\lambda)|\bar q_1(z)\gamma_\mu\gamma_5 q_2 (-z)|0\rangle
 =
 f_{T} m_T^2
  \int\limits_{0}^1 \! du\,e^{i(u-\bar u)pz} \varepsilon_{\mu\nu\alpha\beta} {z^\nu p^\alpha} \epsilon_{(\lambda)}^{*\beta\delta}z_\delta\,{1\over (pz)^2}\, {g}^{\rm BK}_a(u)\,. \nonumber
  \end{eqnarray}
They are related by $g_a(u) =2 \int_0^u {g}^{\rm BK}_a(v) dv\,.$ Note that the variable $t$ used in \cite{Braun:2000cs} is related to $u$ through the relation $t=2u-1$.
Our $g_a$ is defined in the same manner as the LCDA $g_\bot^{(a)}$ in the vector meson case or $g_\bot^{(v)}$ as in the case of the axial-vector meson. This definition is more convenient for studying the relevant Wandzura-Wilczek relation and the helicity projection operator.
}
\begin{eqnarray} \label{eq:chial-even-LCDAs}
 \langle T(P,\lambda)|\bar q_1(z)\gamma_\mu {q_2}(-z)|0\rangle
 &=&
   f_{T} m^2_T
  \int\limits_{0}^1 \! du\,e^{i(u-\bar u)pz} \Bigg\{ p_\mu \frac{\epsilon^{(\lambda)*}_{\alpha\beta}z^\alpha z^\beta}
{(pz)^2} \,\phi_\parallel(u)
\nonumber\\
&+& \frac{\epsilon^{(\lambda)*}_{\perp\mu\alpha} z^\alpha }{pz}\, g_v(u)
 - \frac{1}{2} z_\mu \frac{\epsilon^{(\lambda)*}_{\alpha\beta}z^\alpha z^\beta}{(pz)^3} m_T^2 \,g_3(u)
\Bigg\}\,, \\
 \langle T(P,\lambda)|\bar q_1(z)\gamma_\mu\gamma_5 q_2 (-z)|0\rangle
 &=&
 f_{T} m_T^2
  \int\limits_{0}^1 \! du\,e^{i(u-\bar u)pz} \varepsilon_{\mu\nu\alpha\beta} {z^\nu p^\alpha} \epsilon_{(\lambda)}^{*\beta\delta}z_\delta\,{1\over pz}\, g_a(u)\,,
\end{eqnarray}
and chiral-odd LCDAs to be
\begin{eqnarray}
 \langle T(P,\lambda)|\bar q_1(z)\sigma_{\mu\nu}  q_2(-z)|0\rangle
 &=&
 -i f_{T}^\perp m_T
  \int\limits_{0}^1 \! du\,e^{i(u-\bar u)pz} \Bigg\{\left[\epsilon^{(\lambda)*}_{\perp\mu\alpha} z^\alpha p_\nu
 - \epsilon^{(\lambda)*}_{\perp\nu\alpha} z^\alpha p_\mu\right] \frac{1}{pz} \phi_\perp(u) \nonumber\\
 &+&  (p_\mu z_\nu - p_\nu z_\mu)
 \frac{ m_T^2\epsilon^{(\lambda)*}_{\alpha\beta}z^\alpha z^\beta}{(pz)^3} h_t(u) \nonumber  \\
 &+&
   \frac{1}{2} \left[\epsilon^{(\lambda)*}_{\perp\mu\alpha} z^\alpha z_\nu
 - \epsilon^{(\lambda)*}_{\perp\nu\alpha} z^\alpha z_\mu\right] \frac{m_T^2}{(pz)^2} h_3(u)
  \Bigg\} \,,\\
 \langle T(P,\lambda)|\bar q_1(z)  q_2(-z)|0\rangle
 &=&
-i f_{T}^\perp m_T^3
  \int\limits_{0}^1 \! du\,e^{i(u-\bar u)pz} \frac{\epsilon^{(\lambda)*}_{\alpha\beta}z^\alpha z^\beta}{pz}h_s(u)\,,  \label{eq:chial-odd-LCDAs}
\end{eqnarray}
where $u$ and $\bar u\equiv 1-u$ are the respective momentum fractions carried by $q_1$
and $\bar q_2$ in the tensor meson.
For non-local operators on the lightcone, the path-ordered gauge factor connecting the points $z$ and $-z$ is not explicitly shown here.

In Eqs. (\ref{eq:chial-even-LCDAs})-({\ref{eq:chial-odd-LCDAs}),
$\phi_\parallel, \phi_\perp$ are leading twist-2 LCDAs, and
$g_v, g_a, h_t, h_s$ are twist-3 ones, while $g_3$ and $h_3$, which will not be considered further in this paper, are of twist-4. Throughout the paper we have adopted the conventions
$D_\alpha=\partial_\alpha +ig_s A^a_\alpha \lambda^a/2$ and $\epsilon^{0123}=-1$.

\section{Properties}\label{sec:PRO}

~~~ In SU(3) limit, due to the $G$-parity of the tensor meson, $\phi_\parallel, \phi_\perp, g_v, g_a, h_t, h_s, g_3$ and $h_3$ are antisymmetric under the replacement $u\to 1-u$. Let us take the case of the $a_2$ tensor meson to illustrate the properties of LCDAs. The $G$-parity operator for SU(2) symmetric cases is $\hat{G}=\hat{C}i\tau_2$, where $\hat{C}$ is a charge-conjugation operator and $\tau_2$ the Pauli spinor acting on the isospin space. Because, under the $G$-party transformations
\begin{eqnarray}
      \hat{G} \bar{u}(z)\gamma_\mu d(-z) \hat{G}^\dag
 &=& -\hat{C} \bar{d}(z)\gamma_\mu u(-z) \hat{C}^\dag
  =           \bar{u}(-z)\gamma_\mu d(z) \,, \\
      \hat{G} \bar{u}(z)\gamma_\mu \gamma_5 d(-z) \hat{G}^\dag
  &=&        -\bar{u}(-z)\gamma_\mu \gamma_5 d(z) \,, \\
      \hat{G} \bar{u}(z)\sigma_{\mu\nu} d(-z) \hat{G}^\dag
  &=&         \bar{u}(-z)\sigma_{\mu\nu} d(z) \,, \\
      \hat{G} \bar{u}(z)   d(-z) \hat{G}^\dag
  &=&        -\bar{u}(-z)  d(z) \,,
\end{eqnarray}
for the nonlocal operators and
\begin{equation}
\langle a_2| \hat{G}^\dag = \langle a_2| (-1) \,,
\end{equation}
for the state, we therefore have,
\begin{eqnarray}
 \langle a_2| \bar{u}(z)\gamma_\mu d(-z) |0\rangle
 &=&  \langle a_2| \hat{G}^\dag \hat{G}\bar{u}(z)\gamma_\mu d(-z)\hat{G}^\dag\hat{G} |0\rangle \nonumber\\
 &=& -\langle a_2| \bar{u}(-z)\gamma_\mu  d(z) |0\rangle\,, \\
 \langle a_2| \bar{u}(z)\gamma_\mu \gamma_5 d(-z) |0\rangle
 &=& \langle a_2| \bar{u}(-z)\gamma_\mu \gamma_5 d(z) |0\rangle\,, \\
 \langle a_2| \bar{u}(z)\sigma_{\mu\nu}  d(-z) |0\rangle
 &=& -\langle a_2| \bar{u}(-z)\sigma_{\mu\nu} d(z) |0\rangle\,, \\
 \langle a_2| \bar{u}(z) d(-z) |0\rangle
 &=& \langle a_2| \bar{u}(-z)  d(z) |0\rangle\,.
\end{eqnarray}
Notice that on the right-hand side of equations, the momentum fraction carried by the up quark given by ``$1-u$"  is equivalent to the momentum fraction carried by the anti-down quark on the left-hand side. Therefore, in the SU(2) limit we have $\phi_{\parallel,\perp}(u) =-\phi_{\parallel,\perp}(\bar{u}), g_{v,a,3}(u) = -g_{v,a,3}(\bar{u})$, and $h_{t,s,3}(u) = -h_{t,s,3}(\bar{u})$. This is also true for the isosinglets $f_2(1270)$ and $f'_2(1525)$ which have even $G$-parity quantum numbers and for the isodoublet $K_2^*(1430)$ which is odd under the $G$-parity transformation in SU(3) limit.

Using the QCD equations of motion~\cite{Ball:1998sk,Ball:1998ff}, the two-parton distribution amplitudes $g_v, g_a, h_t$ and $h_s$ can be represented in terms of
$\phi_{\parallel,\perp}$ and three-parton distribution amplitudes. Neglecting the three-parton distribution amplitudes containing gluons and terms proportional to light quark masses, twist-3 LCDAs $g_a,g_v,h_t$ and $h_s$ are related to twist-2 ones through the Wandzura-Wilczek relations:
\begin{eqnarray}\label{eq:WW}
    g_v^{WW}(u) &=& \int\limits_{0}^u dv\, \frac{\phi_\parallel(v)}{\bar v}+
                  \int\limits_{u}^1 dv\, \frac{\phi_\parallel(v)}{v}\,,
\nonumber\\
    g_a^{WW}(u) &=& 2\bar{u}\int\limits_{0}^u dv\, \frac{\phi_\parallel(v)}{\bar v}+
                  2u\int\limits_{u}^1 dv\, \frac{\phi_\parallel(v)}{v}\,,
\nonumber\\
 h_t^{WW}(u) &=& \frac{3}{2} (2u-1)\left(\int\limits_{0}^u dv\, \frac{\phi_\perp(v)}{\bar v} -
                  \int\limits_{u}^1 dv\, \frac{\phi_\perp(v)}{v}\right)\,,
\nonumber\\
    h_s^{WW}(u) &=& 3 \left( \bar{u}\int\limits_{0}^u dv\, \frac{\phi_\perp(v)}{\bar v}+
                  u\int\limits_{u}^1 dv\, \frac{\phi_\perp(v)}{v} \right)\,.
\end{eqnarray}

The leading-twist LCDAs $\phi_{\parallel,\perp}(u,\mu)$ can be expanded as
\begin{eqnarray}\label{eq:conformal-partial-wave-t2}
\phi_{\parallel,\perp}(u,\mu)=6u(1-u)
\sum_{\ell=1}^\infty a_\ell^{\parallel,\perp}(\mu) C^{3/2}_\ell(2u-1),
\end{eqnarray}
where $\mu$ is the normalization scale and the multiplicatively renormalizable
coefficients (or the so-called Gegenbauer moments) are:
\begin{eqnarray}\label{eq:Gegenbauer-moments-t2}
a_\ell^{\parallel,\perp}(\mu) = \frac{2(2\ell+3)}{3(\ell+1)(\ell+2)} \int_0^1
du\, C^{3/2}_\ell (2u-1) \phi_{\parallel,\perp}(u,\mu),
\end{eqnarray}
which vanish with even $\ell$ in the SU(3) limit due to $G$-parity invariance. The Gegenbauer moments $a_l^{\parallel}$ renormalize multiplicatively:
  \begin{equation}
     \left( f^{(\perp)} a_\ell^{\parallel\,(\perp)} \right) (\mu) =
     \left( f^{(\perp)} a_\ell^{\parallel\,(\perp)} \right) (\mu_0)
  \left(\frac{\alpha_s(\mu_0)}{\alpha_s(\mu)}\right)^{-\gamma_{\ell}^{\parallel\,(\perp)}/{b}},
  \label{eq:RGparallel}
   \end{equation}
where $b=(11 N_c -2n_f)/3$ and the one-loop anomalous dimensions are \cite{GW}
  \begin{eqnarray}
  \gamma_{\ell}^\parallel  = C_F
  \left(1-\frac{2}{(\ell+1)(\ell+2)}+4 \sum_{j=2}^{\ell+1} \frac{1}{j}\right),
  \label{eq:1loopandim}
  \end{eqnarray}
\begin{eqnarray}
  \gamma_{\ell}^\perp  = C_F
  \left(1+4 \sum_{j=2}^{\ell+1} \frac{1}{j}\right),
  \label{eq:gamma_perp}
  \end{eqnarray}
with $C_F=(N_c^2-1)/(2N_c )$.

In the present study, the distribution amplitudes are normalized to be
\begin{eqnarray}\label{eq:norm1}
  \int\limits_{0}^{1} du\, (2u-1)\,\phi_{\parallel}(u)
  = \int\limits_{0}^{1} du\, (2u-1)\,\phi_{\perp}(u) = 1 \,.
\end{eqnarray}
Consequently, the first Gegenbauer moments are fixed to be $a_1^\parallel=a_1^\perp={5\over 3}$. Moreover, we have
\begin{eqnarray}\label{eq:norm}
  3 \int\limits_{0}^{1} du\, (2u-1)\,g_a(u) &=&
  \int\limits_{0}^{1} du\, (2u-1)\,g_v(u) = 1\,,\\
  2 \int\limits_{0}^{1} du\, (2u-1)\,h_s(u) &=&
  \int\limits_{0}^{1} du\, (2u-1)\,h_t(u) = 1\,,
\end{eqnarray}
which hold even if the complete leading twist DAs and corrections from the three-parton distribution amplitudes containing gluons are taken into account. The asymptotic wave function is therefore
\begin{eqnarray}\label{eq:phi-as}
    \phi_{\parallel,\perp}^{\rm as}(u) = 30 u(1-u)(2u-1),
\end{eqnarray}
and the corresponding expressions for the twist-3 distributions are
\begin{eqnarray}\label{eq:g-h-as}
 g_v^{\rm as}(u) &=& 5 (2u-1)^3\,,\qquad g_a^{\rm as}(u) =
    10 u(1-u) (2u-1) \,,\nonumber\\
 h_t^{\rm as}(u) &=& \frac{15}{2} (2u-1) (1-6u+6u^2)\,,
 \qquad h_s^{\rm as}(u) = 15 u(1-u) (2u-1)\,.
\end{eqnarray}

\section{Decay constants}\label{sec:DC}

A tensor meson cannot be produced through the usual local $V-A$ and tensor currents, but it can be created through these currents with covariant derivatives (see below). This feature allows us to study its decay constants $f_T$ and $f^\bot_T$.

\subsection{$f_{T}$}
~~~ The decay constant $f_{T}$, which itself involves the Gegenbauer first moment, can be defined through the matrix element of the following operator \footnote{The dimensionless decay constant $f_T$ defined in \cite{Aliev:1981ju,Aliev:2009nn} differs from ours by a factor of $2m_T$. The factor of 2 comes from a different definition of $\stackrel{\leftrightarrow}{D}_\mu$ there.}
\begin{eqnarray} \label{eq:decay-constant}
 \langle T (P,\lambda)|j_{\mu\nu}(0)|0\rangle
 = f_{T} m_T^2 \epsilon^{(\lambda)*}_{\mu\nu},
\end{eqnarray}
where
\begin{equation}
j_{\mu\nu}(0)=\frac{1}{2}
\left( \bar q_1(0)\gamma_\mu i\stackrel{\leftrightarrow}{D}_\nu q_2(0)
     + \bar q_1(0)\gamma_\nu i\stackrel{\leftrightarrow}{D}_\mu q_2(0) \right),
\end{equation}
and $\stackrel {\leftrightarrow}{D}_\mu=\stackrel {\rightarrow}{D}_\mu-\stackrel {\leftarrow}{D}_\mu$ with $\stackrel{\rightarrow}{D}_\mu=\stackrel{\rightarrow}\partial_\mu +ig_s A^a_\alpha \lambda^a/2$ and $\stackrel{\leftarrow}{D}_\mu=\stackrel{\leftarrow}\partial_\mu -ig_s A^a_\alpha \lambda^a/2$.
Its value has been estimated using QCD sum rules for the tensor mesons $f_2(1270)$ \cite{Aliev:1981ju} and $K^*_2(1430)$ \cite{Aliev:2009nn} : \footnote{The decay constants for $f_2(1270)$ and $f'_2(1525)$ had also been estimated in \cite{Bagan:1988ay} using QCD sum rules. The results quoted from \cite{Narison} are:
$f_{f_2(1270)}=(132\sim 184)$ MeV and $f_{f'_2(1525)}=(112\sim 152)$ MeV.}
}
\begin{eqnarray}\label{eq:f-value-1}
    f_{f_2(1270)}(\mu = 1~{\rm GeV}) & \simeq & 0.08\,m_{f_2(1270)}=102~{\rm MeV}\,, \nonumber\\
    f_{K^*_2(1430)}(\mu = 1~{\rm GeV}) & \simeq & (0.10\pm0.01)\,m_{K^*_2(1430)}=(143\pm14)\,{\rm MeV}\,.
\end{eqnarray}
We shall re-analyze the $f_T$ sum rules in the next subsection.

Several authors \cite{Aliev:1981ju,Terazawa:1990es,Suzuki:1993zs} have extracted $f_{f_2(1270)}$ from the measurement of $\Gamma(f_2\to \pi\pi)$
by assuming that the matrix element $\langle\pi^+\pi^-|\Theta_{\mu\nu}|0\rangle$ with $\Theta_{\mu\nu}$ being the energy-momentum tensor is saturated by the $f_2$ meson under the tensor-meson-dominance hypothesis, namely,
\begin{eqnarray}
\langle\pi^+(p)\pi^-(p')|\Theta_{\mu\nu}|0\rangle &\approx& \langle\pi^+(p)\pi^-(p')|f_2\rangle\langle f_2|\Theta_{\mu\nu}|0\rangle \nonumber\\
&=& {f_{f_2}\,g_{f_2\pi\pi}m_{f_2}\over (p+p')^2-m^2_{f_2}}(p-p')_\mu (p-p')_\nu,
\end{eqnarray}
where $g_{f_2\pi\pi}$ is the coupling constant defined by
\begin{eqnarray}
\langle\pi^+(p)\pi^-(p')|f_2\rangle ={g_{f_2\pi\pi}\over m_{f_2}}\epsilon^{\mu\nu}_{(\lambda)}(p-p')_\mu (p-p')_\nu\,.
\end{eqnarray}
The decay rate reads
\begin{eqnarray}
\Gamma(f_2\to \pi^+\pi^-) ={4\over 15\pi m^2_{f_2}}\,\left({g_{f_2\pi\pi}\over m_{f_2}}\right)^2p_c^5,
\end{eqnarray}
with $p_c$ being the center-of-mass momentum of the pion.
From the measured width $\Gamma(f_2\to \pi\pi)=(156.9^{+4.0}_{-1.2})$ MeV \cite{PDG} and the normalization condition $\langle \pi(p)|\Theta_{00}|\pi(p)\rangle=2m_\pi^2$ \cite{Terazawa:1990es}, we obtain
\begin{eqnarray}
    f_{f_2(1270)} \simeq (0.085\pm 0.001) m_{f_2(1270)}=(108\pm1)\,{\rm MeV}\,,
\end{eqnarray}
which is in agreement with \cite{Suzuki:1993zs}. By the same token, if the matrix element $\langle K^+K^-|\Theta_{\mu\nu}|0\rangle$ is assumed to be saturated by $f'_2(1525)$ which is $s\bar s$ dominated, we will have
\begin{eqnarray}
    f_{f'_2(1525)} \simeq (0.089\pm 0.003) m_{f'_2(1525)}=(136\pm5)\,{\rm MeV}\,,
\end{eqnarray}
where use of the experimental value $\Gamma(f'_2\to K\bar K)=(65^{+5}_{-4})$ MeV \cite{PDG} has been made.

\subsection{$f_{T}^\perp$}

~~~ Using the QCD sum rule technique, we proceed to estimate the value of $f_T^\perp$ \cite{SVZ}. To determine the magnitude and the relative sign of $f_{T}^\perp$ with respect to $f_T$, we consider the non-diagonal two-point correlation function,
\begin{eqnarray}\label{eq:decay-const-2pt}
i (2\pi)^4 \delta^4(q-p)\Pi_{\mu\nu\delta\alpha\beta} (q) &=&
 i^2\int d^4x d^4 y e^{i(qx-py)}
 \langle 0|{\rm T} [j^{\perp\dag}_{\mu\nu\delta}(x) \, j_{\alpha\beta}(y)]|0\rangle\,,
\end{eqnarray}
with
\begin{eqnarray}\label{eq:Pi}
\Pi_{\mu\nu\delta\alpha\beta} (q)
 &=& \frac{i}{2}\left[(g_{\alpha\mu} g_{\beta\delta} + g_{\alpha\delta} g_{\beta\mu}) q_\nu
               - (g_{\alpha\nu} g_{\beta\delta} + g_{\alpha\delta} g_{\beta\nu}) q_\mu \right]\Pi(q^2) + \dots  \,.
\end{eqnarray}
The interpolating current $j^{\perp\dag}_{\mu\nu\delta}(0)=\bar q_2(0) \sigma_{\mu\nu} i\stackrel{\leftrightarrow}{D}_{_\delta}(0) q_1(0)$ satisfies the relation
 \begin{eqnarray}
    \langle 0 |j^{\perp\dag}_{\mu\nu\delta}(0) |T(P,\lambda) \rangle
 =i  f_{T}^\perp m_T
 (\epsilon_{\mu\delta}^{(\lambda)*} P_\nu-\epsilon_{\nu\delta}^{(\lambda)*} P_\mu ) .
  \label{eq:tensor-decayconstant}
 \end{eqnarray}
Here we are only interested in the Lorentz invariant constant $\Pi(q^2)$ which receives the contribution from tensor mesons but not from vector or scalar mesons.

\begin{figure}[t]
\begin{center}
\includegraphics[width=12cm]{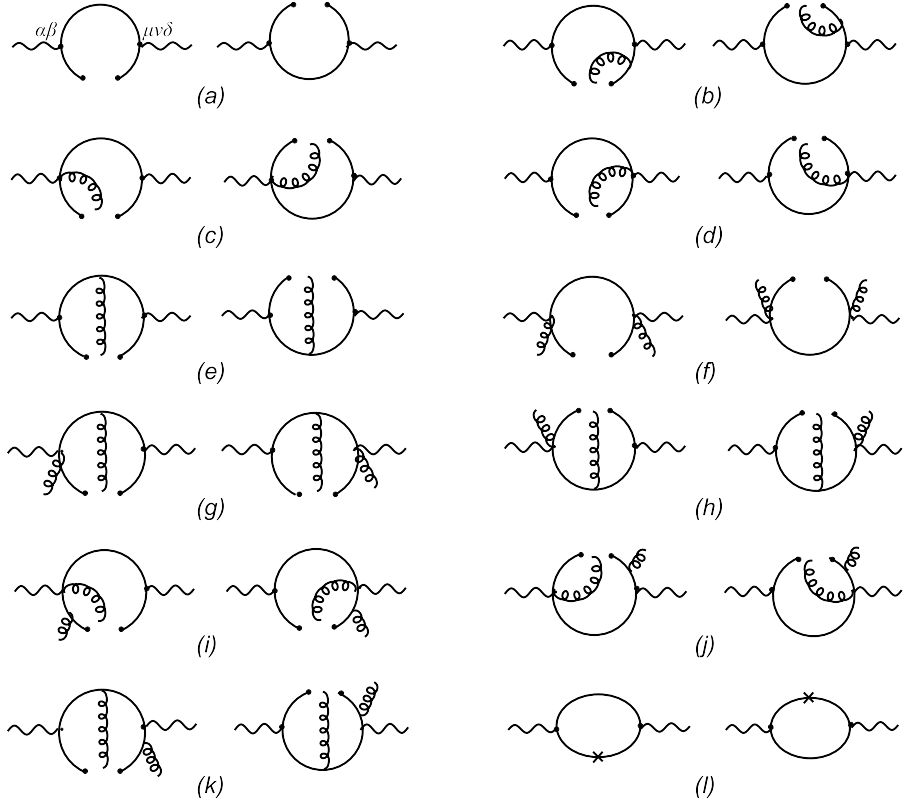}
\vspace{0.0cm}
\caption{Diagrams contributing to the OPE expansion of the two-point correlation function $\Pi(q^2)$ defined in Eqs. (\ref{eq:decay-const-2pt-2}) and (\ref{eq:Pi}). Diagrams (c), (f) and the left diagrams of (g), (h) , (i) and (j) involving a soft gluon emitted from the left vertex do not contribute to $\Pi_{\mu\nu\delta\alpha\beta}(q)$, while both diagrams in (e) also make no conributions to the invariant structure of $\Pi(q^2)$. The cross signs in Fig. (l) denote a mass insertion.} \label{fig:OPE} \end{center}
\end{figure}

To simply the calculation of $\Pi(q^2)$, we will apply the translation transformation to the current $j_{\alpha\beta}(y)$
\begin{eqnarray}
j_{\alpha\beta}(y)=e^{i\hat{P}(y-z)} j_{\alpha\beta}(z) e^{-i\hat{P}(y-z)} ,
\end{eqnarray}
where $\hat{P}$ is a translation operator, and then recast Eq. (\ref{eq:decay-const-2pt}) to
\begin{eqnarray}\label{eq:decay-const-2pt-2}
i (2\pi)^4 \delta^4(q-p)\Pi_{\mu\nu\delta\alpha\beta} (q) &=&
 i^2 (2\pi)^4 \delta^4(q-p) \int d^4 x^\prime e^{iq(x^\prime -z)}
 \langle 0|{\rm T} [j^{\perp\dag}_{\mu\nu\delta}(x^\prime) \, j_{\alpha\beta}(z)]|0\rangle \vert_{z\to 0}\,. \nonumber \\
\end{eqnarray}
The covariant derivative $\stackrel{\leftrightarrow}{D}_{\beta}(z)$ in $\bar q_1(0)\gamma_\alpha i\stackrel{\leftrightarrow}{D}_\beta q_2(z)$
then becomes
\begin{eqnarray}
   \stackrel{\leftrightarrow}{D}_{\beta}(z)
 &=& \frac{\stackrel{\rightarrow}{\partial}}{\partial z^\beta}
 -\frac{\stackrel{\leftarrow}{\partial}}{\partial z^\beta}
 +  i g_s \lambda^a A^a_\beta(z) \nonumber \\
 &=& \frac{\stackrel{\rightarrow}{\partial}}{\partial z^\beta}
 -\frac{\stackrel{\leftarrow}{\partial}}{\partial z^\beta}
 + {1\over 2}i g_s \lambda^a z^\lambda G^a_{\lambda\beta}(z) +\cdots \,,
\end{eqnarray}
in the fixed-point gauge (or the so-called Schwinger-Fock gauge) \cite{SVZ}
\begin{equation}\label{eq:fix-point-gauge}
z^\beta A^a_\beta(z)=0~~{\rm with}~~A_\beta^a (z) =\int_0^1 dt\, t z^\lambda G_{\lambda\beta}(tz)\,.
\end{equation}
Consequently, $\stackrel{\leftrightarrow}{D}_{\beta}(z)$ is reduced to the usual derivative
$\stackrel{\rightarrow}{\partial}/\partial z^\beta- \stackrel{\leftarrow}{\partial}/\partial z^\beta$ in the $z \to 0$ limit and hence the contributions from the diagrams in Fig.~\ref{fig:OPE} with the soft gluons emerging from the left vertex vanish. Likewise, the uses of the translation transformation for $j^{\perp\dag}_{\mu\nu\delta}(x)$
\begin{eqnarray}
j^{\perp\dag}_{\mu\nu\delta}(x)=e^{i\hat{P}(x-z)} j^{\perp\dag}_{\mu\nu\delta}(z) e^{-i\hat{P}(x-z)} ,
\end{eqnarray}
and the corresponding relation for the non-diagonal two-point correlation function
\begin{eqnarray}\label{eq:decay-const-2pt-3}
i (2\pi)^4 \delta^4(q-p)\Pi_{\mu\nu\delta\alpha\beta} (q) &=&
 i^2 (2\pi)^4 \delta^4(q-p) \int d^4 y^\prime e^{-iq(y^\prime -z)}
 \langle 0|{\rm T} [j^{\perp\dag}_{\mu\nu\delta}(z) \, j_{\alpha\beta}(y^\prime)]|0\rangle \vert_{z\to 0}\, \nonumber\\
\end{eqnarray}
will imply that the diagrams with the soft gluons emerging from the right vertex vanish. Note that one can apply either Eq. (\ref{eq:decay-const-2pt-2}) or (\ref{eq:decay-const-2pt-3}) to compute the two-point correlation function $\Pi(q^2)$; the results should be the same. In this work we shall use the former to evaluate the operator-product expansion (OPE) of $\Pi(q^2)$.

%

The resulting $\Pi^{\rm OPE}$, which is the OPE
result of $\Pi(q^2)$ up to dimension-7 at the quark-gluon level (see Fig. \ref{fig:OPE}), reads
\begin{eqnarray}\label{eq:OPE}
   \Pi^{\rm OPE} (q^2)
  &\cong&  -\frac{1}{4} (\langle \bar q_1 q_1\rangle + \langle \bar q_2 q_2\rangle)
           -\frac{7}{12q^2}
             (\langle \bar q_1 g_s \sigma G q_1\rangle + \langle \bar q_2 g_s \sigma G q_2\rangle) \nonumber\\
  & + & \frac{1}{96 q^4}
        (\langle \bar q_1 q_1\rangle + \langle \bar q_2 q_2\rangle)\langle g_s^2 G^2\rangle
 -\frac{m_{q_1} +m_{q_2}}{32\pi^2} q^2 [\ln (-{q^2/ \mu^2})-1],
\end{eqnarray}
where $\mu$ is the renormalization scale, the first term on the right-hand side of the above equation arises from Fig. 1({\it a}), the second from 1({\it b}) and 1({\it d}), the third from 1({\it k}) and the right diagrams of 1({\it g}), 1({\it h}), 1({\it i}) and 1({\it j}), and the fourth from 1({\it l}). We have adopted the shorthand notation $\langle \dots \rangle \equiv \langle 0|: \dots :|0 \rangle$ for vacuum condensates. In the calculation the ultraviolet divergence is regularized by using the modified minimal substraction ($\overline{\rm MS}$) scheme.

To suppress the non-resonant background in the sum rules, we take into account the dispersion relation with
a subtraction. This method was first introduced in Refs.~\cite{Balitsky:1982dt,Braun:1985ah}. Considering $\tilde
\Pi(q^2)= \Pi(q^2)- \Pi^{\rm pert}(q^2)$, which is finite in the limit $-q^2\to \infty$,  we then
get
\begin{eqnarray}\label{eq:ope-2}
\tilde\Pi (q^2) &=& \tilde
\Pi(0)+\frac{q^2}{\pi}\int_0^\infty\frac{ds}{s(s-q^2)}
 [\rho_{phys} (s) - {\rm Im}\Pi^{\rm pert}(s)] \,,
\end{eqnarray}
where $\rho_{phys}$ and $\rho_{tensor}$ are the total physical and lowest-lying tensor meson spectral
densities, respectively, which can be modeled as
 \begin{eqnarray}
 \rho_{phys}(s)&=& \rho_{tensor} (s) +\theta(s-s_0){\rm Im}\Pi^{\rm pert}(s) \nonumber \\
 &=&f_T f_T^\perp m_T^3 \pi \delta(s-m_T^2) +\theta(s-s_0){\rm Im}\Pi^{\rm pert}(s)\,.
 \end{eqnarray}
Here $s_0$ is the excited threshold and the imaginary part of $\Pi^{\rm pert}(s)$ is
\begin{equation}\label{imPi-OPE}
{\rm Im} \Pi^{\rm pert}(s)=\frac{m_{q_1} +m_{q_2}}{32\pi^2}s.
\end{equation}
Taking the limit $-q^2\to \infty$ in Eq. (\ref{eq:ope-2}), we obtain the following relation:
\begin{eqnarray}\label{eq:Pi(0)}
\tilde\Pi (0) &=& -\frac{1}{4} (\langle \bar q_1q_1\rangle + \langle \bar q_2q_2\rangle) +
 f_T f_T^\perp m_T   - \frac{1}{\pi}
 \int_0^{s_0}\frac{ds}{s}{\rm Im}\Pi^{\rm pert}(s) \,.
\end{eqnarray}
After performing the Borel transformation \cite{SVZ,Yang:2007zt} and taking into account scale-dependence of each quantities, we arrive at the sum rule:
\begin{eqnarray}\label{eq:SR2}
 f_T f_T^\perp &\cong&  \frac{1}{(e^{-m_T^2/M^2}-1)m_T }
   \left[-\frac{7}{12} \frac{\langle \bar q_1 g_s \sigma G q_1\rangle
 + \langle \bar q_2 g_s \sigma G q_2 \rangle}{ M^2}  \right. \nonumber\\
 & & - \frac{\pi}{48 M^4}
        (\langle \bar q_1 q_1\rangle + \langle \bar q_2 q_2\rangle)\langle \alpha_s G^2\rangle
         \left. +\frac{m_{q_1} +m_{q_2}}{32\pi^2}M^2 \left( 1-e^{-s_0/M^2} -\frac{s_0}{M^2} \right)
   \right] .
\end{eqnarray}
In the numerical analysis, we shall use the following  input parameters at the scale 1~GeV \cite{Yang:2007zt}:
\begin{eqnarray}
\begin{array}{lcl}
 \alpha_s(1~{\rm GeV})=0.497\pm 0.005 \,, & &  m_s(1~{\rm GeV})=(140\pm 20)~ {\rm MeV}, \\
 \langle \bar uu \rangle \cong \langle \bar dd \rangle =-(0.240\pm 0.010)^3~ {\rm GeV}^3 \,,
  &   & \langle \bar ss \rangle = (0.8\pm 0.1) \langle \bar uu \rangle \,, \\
 \langle g_s \bar u\sigma Gu \rangle \cong
  \langle g_s\bar d\sigma Gd \rangle =-(0.8\pm 0.1)\,{\rm GeV}^2\langle \bar uu \rangle, & &
  \langle g_s \bar s\sigma Gs \rangle = (0.8\pm 0.1) \langle g_s\bar u\sigma Gu \rangle, \\
 \langle \alpha_s G_{\mu\nu}^a G^{a\mu\nu} \rangle =(0.474\pm 0.120)\ {\rm GeV}^4/(4\pi)\,. &  &
\end{array}\label{eq:parameters}
 \end{eqnarray}
The masses of $u$- and $d$-quarks can be numerically neglected.
For the separate determination of $f_T$ and $f_T^\perp$,
we next proceed to re-analyze the $f_T$ sum rule which is given by \cite{Aliev:1981ju,Aliev:2009nn}
\begin{eqnarray}\label{eq:SR3}
 & &f_T^2 e^{-m_T^2/M^2} \cong  \frac{1}{m_T^4}
 \left\{ \frac{3}{20\pi^2}M^6 \left[1-\left( 1+\frac{s_0}{M^2}+ \frac{s_0^2}{2M^4}\right)e^{-s_0/M^2}\right]
 - \frac{2M^2}{9\pi} \langle\alpha_s^2 G^2\rangle  \right.
        \nonumber\\
 & & ~~~~~~~ \left. + \frac{32\pi\alpha_s}{9} \langle \bar q_1 q_1\rangle \langle \bar q_2 q_2\rangle
 + \frac{m_{q_2}\langle \bar q_1 g_s \sigma G q_1\rangle
 +       m_{q_1}\langle \bar q_2 g_s \sigma G q_2 \rangle}{ 6} \right\}
  \,.
\end{eqnarray}
For the sum rule calculation, the decay constants and parameters are evaluated at $\mu=1~{\rm GeV}$. Changing the scale within the range $\mu^2=(1-2)\,{\rm GeV}^2$ does not cause any noticeable effect, provided that the decay constants are also rescaled according to the renormalization group equation.
Applying the differential operator $M^4\partial/ \partial M^2$ to the above equation, we  obtain the mass sum rule for the tensor meson, from which we can determine (i) the excited threshold $s_0$ and (ii) the working Borel window $M^2$ where the resulting tensor mass is well stable and in agreement with the data. However, we note that the contribution originating from modelling higher resonances defined by
\begin{eqnarray}\label{eq:resonance}
 & & \left[ \frac{3}{20\pi^2}M^6 \left( 1+\frac{s_0}{M^2}+ \frac{s_0^2}{2M^4}\right)e^{-s_0/M^2} \right] \Bigg{/}
 \left[ \frac{3}{20\pi^2}M^6
 - \frac{2M^2}{9\pi} \langle\alpha_s^2 G^2\rangle  \right.
        \nonumber\\
 & & ~~~~~~~ \left. + \frac{32\pi\alpha_s}{9} \langle \bar q_1 q_1\rangle \langle \bar q_2 q_2\rangle
 + \frac{m_{q_2}\langle \bar q_1 g_s \sigma G q_1\rangle
 +       m_{q_1}\langle \bar q_2 g_s \sigma G q_2 \rangle}{ 6} \right]
  \,
\end{eqnarray}
is about 60\% for $M^2=1.0$ GeV$^2$ and 80\% for $M^2=1.6$ GeV$^2$. The higher resonance corrections may be a bit too large but still controllable. On the other hand, when $M^2> 1.0$ GeV$^2$, the highest OPE term at the quark-gluon level is no more than 8\% which is relatively small.

\begin{table}[t]
\caption[]{Sum rule results for the decay constants $f_T$ and $f_T^\bot$ of various tensor mesons at the scale $\mu=1$~GeV. The results for the excited threshold $s_0$, masses of the tensor mesons, Borel windows $M^2$ (in units of GeV$^2$), and $f_T$ are obtained from Eq. (\ref{eq:SR3}), $f_T f_T^\perp$ from Eq. (\ref{eq:SR2}) and $f_T^\perp$ from the combination of $f_T$ and $f_Tf_T^\perp$. The error for $f_T$ is due mainly to the uncertainties in vacuum condensates, while the first error in $f_T f_T^\bot$ arises from the Borel mass and the second error from the rest of other input parameters.
}
\label{tab:decay-constant}
\renewcommand{\arraystretch}{1.5}
\addtolength{\arraycolsep}{0pt}
$$
\begin{array}{|c|c|c|c|c|c|c|}\hline
  {\rm State} & s_0\ ({\rm GeV}^2) &  {\rm Range\ of}~ M^2 & {\rm Mass\ (GeV)}
&  f_T\ ({\rm MeV}) & f_{T} f_T^\perp\ ({\rm MeV}^2) &  f_T^\perp\ ({\rm MeV})
 \\ \hline
\begin{array}{c} f_2(1270)   \\ f_2^\prime(1525) \\ a_2(1320) \\ K_{2}^*(1430) \end{array}&
\begin{array}{c} 2.53 \\ 3.49 \\  2.70 \\ 3.13   \end{array} &
\begin{array}{c} (1.0,1.4) \\ (1.3,1.7)\\  (1.0,1.4) \\ (1.2,1.6)   \end{array} &
\begin{array}{c} 1.27\pm0.01 \\ 1.52\pm0.02 \\ 1.31\pm0.01 \\ 1.43\pm0.01   \end{array} &
\begin{array}{c} 102\pm 6 \\ 126\pm4 \\  107\pm6 \\ 118\pm5   \end{array}&
\begin{array}{c} 11900\pm700\pm1600 \\ 8200\pm300\pm1100 \\  11200\pm600\pm1500 \\ 9100\pm500\pm1200   \end{array} &
\begin{array}{c} 117\pm 25 \\ 65\pm12 \\  105\pm21 \\ 77\pm14   \end{array}
\\ \hline
\end{array}
$$
\end{table}
%

We then estimate $f_T$ and $f_T^\perp$ from Eqs. (\ref{eq:SR3}) and (\ref{eq:SR2}), respectively. All the numerical results are collected in Table \ref{tab:decay-constant}. Here we have assumed that the obtained $s_0$ and corresponding Borel window are applicable to both $f_T$ and $f_T f_T^\perp$ sum rules.  The theoretical
errors are due to the variation of the Borel mass, quark masses, and vacuum condensates, which are then added in quadrature. For simplicity, we do not take into account the uncertainty in $s_0$. 
In the analysis, we have neglected the possible mixture of the quark and gluon currents for $f_2(1270)$ and $f_2^\prime(1525)$ mesons. As noticed in the Introduction, we assume that $f_2(1270)$ is a $(u\bar u+d\bar d)/\sqrt{2}$ state, while $f'_2(1525)$ is predominantly made of $s\bar s$. Our results are in good agreement with \cite{Aliev:1981ju} for $f_{f_2(1270)}$, but smaller than that of \cite{Aliev:2009nn} for $f_{K_2^*(1430)}$. We should note that our $f_T f_T^\perp$ is obtained from the non-diagonal sum rule and hence it is insensitive to $s_0$. For the non-diagonal sum rule, one possible error may arise from the radiative corrections, which are at about  10\% level for each OPE term and partly contribute to higher resonances, and can be lumped into the uncertainties of the input parameters given in Eq. (\ref{eq:parameters}).

\section{Conclusion}\label{sec:sum}

~~~ We have systematically studied the two-parton light-cone distribution amplitudes for $1\,^3P_2$ nonet tensor mesons. The light-cone distribution amplitudes can be presented by using QCD conformal partial wave expansion. We have obtained the asymptotic two-parton distribution amplitudes of twist-2 and twist-3. The relevant decay constants have been estimated using the QCD sum rule techniques. We have also studied the decay constants for $f_2(1270)$ and $f_2^\prime(1525)$ based on the hypothesis of tensor meson dominance together with the data of $\Gamma(f_2\to \pi\pi)$ and $\Gamma(f'_2\to K\bar K)$. The results are in accordance with the sum rule predictions.

\subsection*{Acknowledgements}
One of us (H.Y.C.) wishes to thank C.N. Yang
Institute for Theoretical Physics at SUNY Stony Brook for its
hospitality. This work was supported in part by the National Center for Theoretical Sciences and the National Science Council of R.O.C. under Grant Nos. NSC96-2112-M-033-004-MY3,
NSC97-2112-M-008-002-MY3 and NSC99-2112-M-003-005-MY3.

\end{document}